\documentclass[prd,twocolumn,preprintnumbers,showpacs]{revtex4}
\usepackage{amsmath,amssymb,bm,graphicx}

\newcommand\grad{\bm{\nabla}}
\newcommand\p{{\bm{p}}}
\newcommand\q{{\bm{q}}}
\renewcommand\k{{\bm{k}}}
\renewcommand\a{{\bm{a}}}
\renewcommand\r{{\bm{r}}}
\newcommand\Lag{\mathcal{L}}

\begin{document}
\preprint{INT-PUB 07-20}

\title{Renormalization group analysis of resonantly interacting anyons}
\author{Yusuke~Nishida}
%\email{nishida@phys.washington.edu}
%\homepage{http://tkynt2.phys.s.u-tokyo.ac.jp/~nishida/}
\affiliation{Institute for Nuclear Theory, University of Washington,
             Seattle, Washington 98195-1550, USA}

\begin{abstract}
 We formulate a field theory for resonantly interacting anyons, that
 enables us to perform a perturbative calculation near the fermionic
 limit.  We derive renormalization group equations for three-body and
 four-body couplings at one-loop order.  In addition to two fixed
 points, we find a limit cycle behavior in the four-body coupling, which
 implies an infinite set of bound states in the four-anyon system.
\end{abstract}

\date{August 2007}
%05.30.Pr  Fractional statistics systems (anyons, etc.)
%11.10.Hi  Renormalization group evolution of parameters
%21.45.+v  Few-body systems

\maketitle

\section{Introduction}
Physics in two-dimensional space has intriguing features that cannot be
found in three dimensions.  One such example is anyons, quantum
particles having fractional
statistics~\cite{Leinaas:1977fm,Wilczek:1981du}.  The statistics of
anyons interpolates continuously between usual fermion and boson cases.
Anyons are not only theoretically interesting but also can be realized
as quasiparticles in the fractional quantum Hall effect~\cite{FQHE}.
Recently it is argued that anyonic quasiparticles may be created using
rotating Bose-Einstein condensates~\cite{BEC} or cold atoms confined in
an optical lattice~\cite{lattice}.  Furthermore anyons are attracting
renewed interest in the context of quantum
computation~\cite{Kitaev:1997wr}.

In this article, we investigate nonrelativistic anyons interacting via a
contact interaction.  Motivated by recent experimental realization of
fermions and bosons at infinite scattering length, we focus on the case
where the two-body interaction is tuned to the resonance at threshold,
i.\,e., two anyons form a bound state with zero binding energy.  Because
of its scale invariance, such a system should be described by
nonrelativistic conformal field theories (CFT).  A field-theoretical
description of resonantly interacting anyons known so far allows for
perturbative calculations only near the bosonic
limit~\cite{Bergman:1993kq,AmelinoCamelia:1994we}~\footnote{Although it
is not mentioned explicitly, the attractive contact interaction with the
critical coupling in Ref.~\cite{Bergman:1993kq} corresponds to the
two-body resonant interaction.}.

Here we propose an alternative description that becomes perturbative
near the fermionic limit and study a few-body problem of resonantly
interacting anyons.  For more than two anyons, additional three-body and
four-body contact interactions turn out to be necessary to ensure the
renormalizability of the theory.  In the renormalization group (RG)
equations of three-body and four-body couplings, we will find two
nontrivial fixed points and a limit cycle behavior of the four-body
coupling.  Our finding implies the existence of an infinite set of
discrete bound states in the four-anyon system, where the ratio of two
successive binding energies is given by
\begin{equation}
 \frac{E_{n+1}}{E_n}\to\exp\!\left(-\frac{\pi}{3|\alpha|}\right)
\end{equation}
near the fermionic limit $|\alpha|\ll1$.

\section{Anyons in quantum mechanics}
Before going on to the field-theoretical formulation of anyons with the
contact interaction, it is instructive to review two-anyon problem in
quantum mechanics.  Anyons are characterized by a phase
$e^{i\pi(\alpha-1)}$ acquired by the wave function under the exchange of
two anyons.  Without loss of generality, the statistics parameter
$\alpha$ is assumed to be in the interval $[-1,1]$.  For the later
convenience, we chose $\alpha$ so that $\alpha=0$ corresponds to
fermions and $|\alpha|=1$ corresponds to bosons.

In the center of mass frame, the two-body wave function can be expressed
as $\Psi_\alpha(\r)=e^{i\alpha\varphi}\Phi(r,\varphi)$, where $r$ and
$\varphi$ are relative distance and angle between two anyons.
$\Phi(r,\varphi)$ is a fermionic wave function which is antisymmetric
under $\varphi\to\varphi+\pi$.  Since $\Psi_\alpha(\r)$ obeys a free
Schr\"odinger equation, the wave function expanded in partial waves
$\Phi(r,\varphi)=\sum_le^{il\varphi}R_l(r)$ satisfies (below $\hbar=1$
and anyon mass $m=1$)
\begin{equation}\label{eq:schrodinger}
 \left[\frac{d^2}{dr^2}+\frac1r\frac{d}{dr}
  -\frac{\left(l+\alpha\right)^2}{r^2}+E\right]R_l(r)=0
\end{equation}
for $r\neq0$, where the odd integer $l$ is an orbital angular momentum.

The contact interaction between anyons can be introduced by allowing
a singular solution that behaves as $R_l(r)\sim r^{-|l+\alpha|}$ at
origin $r\sim0$, i.\,e., when the positions of two anyons
coincide~\cite{Grundberg:1990qn,Manuel:1990in,Bourdeau:1991fw}.
Because of the square integrability of the wave function, the contact
interaction is available only for $l=-1$ when $\alpha>0$ or for $l=1$
when $\alpha<0$.  In this channel, the contact interaction is equivalent
to imposing the following boundary condition on the wave function at
origin: 
\begin{equation}\label{eq:wavefunction}
 R_{\mp1}(r) \to \frac1{r^{1-|\alpha|}}-\mathrm{sgn}(a)
  \left(\frac{r}{a^2}\right)^{1-|\alpha|}+O(r^{1+|\alpha|}),
\end{equation}
where $a$ is a real parameter analogous to the scattering length in
three spatial dimensions.  There are two scale-independent boundary
conditions; $a\to0$ and $a\to\infty$.  The former corresponds to the
usual non-colliding limit.  Since the binding energy is given
by~\cite{Manuel:1990in}
\begin{equation}
 E_{\mathrm b} = -\frac4{a^2}\left[\frac{\Gamma(2-|\alpha|)}
	{\Gamma(|\alpha|)}\right]^{\frac1{1-|\alpha|}}
\end{equation}
when $a$ is positive, the resonant limit corresponds to $a\to\infty$.

An interesting observation is that the normalization integral of the
wave function (\ref{eq:wavefunction}) is logarithmically divergent at
$r\to0$ in the fermionic limit $|\alpha|\to0$.  Therefore, if the
statistics is close to fermion, the two-body wave function is
concentrated near the origin, and the anyon pair forms a pointlike
dimer.  The same situation has been observed in two-component fermions
at infinite scattering length near four spatial
dimensions~\cite{Nussinov06}.  This observation has led to a field
theory composed of weakly coupled fermions and
dimers~\cite{Nishida:2006br,Nikolic-Sachdev07}.  Similarly, as we shall
see below, it is possible to formulate a field theory describing
resonantly interacting anyons, which allows for a perturbative analysis
near the fermionic limit.

\section{Field-theoretical formulation}
The field-theoretical representation of anyons is provided by a
nonrelativistic field $\psi$ minimally coupled to a Chern-Simons (CS)
gauge field $(a_0,\a)$~\cite{Hagen:1984mj,Jackiw:1990mb}:
\begin{equation}\label{eq:L_psi}
 \begin{split}
  \Lag_\psi &= \frac1{4\pi\alpha}\partial_t\a\times\a
  -\frac1{2\pi\alpha}a_0\grad\times\a
  -\frac1{2\zeta}\left(\grad\cdot\a\right)^2 \\
  &\quad+\psi^*\left(i\partial_t-a_0+\mu\right)\psi
  -\frac12\left|\left(\grad-i\a\right)\psi\right|^2.
 \end{split}
\end{equation}
In our definition of $\alpha$, $\psi$ is a fermionic field.  $\mu$ is
the anyon chemical potential and here we consider the system at zero
density $\mu=0$.  We shall work in the Coulomb gauge $\zeta=0$.

Motivated by the above argument, it is natural to add the following
terms involving the dimer field $\phi$ to (\ref{eq:L_psi}):
\begin{align}
 \Lag_\phi &= \phi^*\left(i\partial_t-2a_0-\epsilon_0\right)\phi
  -\frac14\left|\left(\grad-2i\a\right)\phi\right|^2 \\
  &\quad+g\,\phi^*\psi\left(-i\partial_x\pm\partial_y\right)\psi
  +g\,\psi^*\left(-i\partial_x\mp\partial_y\right)\psi^*\phi. \notag
\end{align}
$\phi$ is a bosonic field with mass $2m$ and has the orbital angular
momentum $l=\mp1$.  The upper sign corresponds to $\alpha>0$ and the
lower sign corresponds to $\alpha<0$ throughout this article.  $\phi$
also couples to the CS gauge field in a gauge invariant way.  $g$ is a
dimensionless coupling between the dimer and a pair of anyons in
$p$-wave.  $\epsilon_0$ is a cutoff-dependent bare ``mass gap'', which
is fine-tuned so that the $\phi$ propagator has a pole at $p_0=\p=0$,
i.\,e.,
\begin{equation}
 \epsilon_0=2g^2\int^\Lambda\!\frac{d\k}{(2\pi)^2}+O(\alpha g^2).
\end{equation}
This condition is equivalent to the fine-tuning to the resonant limit
(zero binding energy).

The other marginal operators one can add to the Lagrangian density are
three-body and four-body contact interactions:
\begin{equation}
 \Lag_\delta = v_3\phi^*\psi^*\psi\phi+v_4\phi^*\phi^*\phi\phi.
\end{equation}
Note that terms as $\grad{\,\times\,}\a\,\psi^*\psi$ and
$\grad{\,\times\,}\a\,\phi^*\phi$ are also allowed, but they can be
absorbed by the redefinition of $v_3$ and $v_4$ if we use the CS Gauss
law
\begin{equation}\label{eq:gauss}
 \grad\times\a=-2\pi\alpha\left(\psi^*\psi+2\phi^*\phi\right).
\end{equation}
Thus, the most general renormalizable Lagrangian density composed of
$\psi$ and $\phi$ becomes $\Lag=\Lag_\psi+\Lag_\phi+\Lag_\delta$.  
First, we concentrate on the two-body sector, i.\,e., a renormalization
of the two-body coupling $g$.

\section{Two-anyon physics}
\begin{figure}[tp]
 \includegraphics[width=0.48332\textwidth,clip]{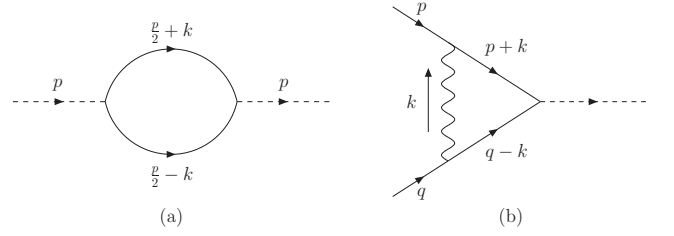}
 \caption{(a) One-loop self-energy diagram to renormalize the wave
 function of $\phi$.  (b) One-loop vertex diagram to renormalize the
 two-body coupling $g$.  The solid, dashed, and wavy lines represent
 propagators of $\psi$, $\phi$, and CS gauge fields, respectively.
 \label{fig:two-anyon}}
\end{figure}

The renormalization of the theory can be performed in the standard
way.  There is a one-loop self-energy diagram for $\phi$ that is
logarithmically divergent [Fig.~\ref{fig:two-anyon}(a)].
Integrating out modes in the momentum shell $e^{-s}\Lambda<k<\Lambda$,
we obtain
\begin{equation}
 \Sigma(p) = -\frac{g^2}{\pi}\left(p_0-\frac{\p^2}4\right) 
  \ln\frac\Lambda{e^{-s}\Lambda},
\end{equation}
which corresponds to the wave-function renormalization of $\phi$;
$Z_\phi=1-(g^2/\pi)s$.  The vertices of $\phi$ with $a_0$ or $\a$ are
also renormalized in the same way.  Thus, the anomalous dimension of
$\phi$ is found to be
\begin{equation}
 \gamma_\phi = -\frac12 \frac{\partial\ln Z_\phi}{\partial s}
  = \frac{g^2}{2\pi}.
\end{equation}

The vertex $g\,\phi^*\psi\left(-i\partial_x\pm\partial_y\right)\psi$ is
renormalized by the logarithmically divergent one-loop diagram in
Fig.~\ref{fig:two-anyon}(b), which is evaluated as
\begin{equation}
 i|\alpha|g\left[\left(\p-\q\right)_x
  \pm i\left(\p-\q\right)_y\right]\ln\frac\Lambda{e^{-s}\Lambda}.
\end{equation}
As a result, a RG equation that governs the running of $g$ is given by
\begin{equation}
 \frac{dg}{ds} = -g\gamma_\phi+|\alpha|g
  = -\frac{g^3}{2\pi}+|\alpha|g.
\end{equation}
There is a nontrivial fixed point located at
\begin{equation}\label{eq:g*}
 {g^*}^2 = 2\pi|\alpha|.
\end{equation}
At this fixed point, the theory is a nonrelativistic CFT describing
anyons at the two-body resonance.  Since ${g^*}^2\sim\alpha\ll1$ in the
fermionic limit, one can perform a perturbative analysis in terms of
$\alpha$.

\begin{figure}[tp]
 \includegraphics[width=0.46\textwidth,clip]{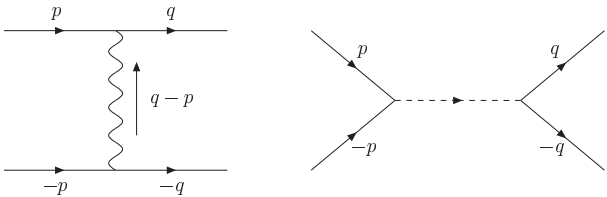}
 \caption{Tree diagrams for two-anyon scattering.
 \label{fig:aharonov-bohm}}
\end{figure}

In order to confirm the connection of physics described by this fixed
point with resonantly interacting anyons, we compute two physical
quantities at $g=g^*$ using perturbative expansions in terms of $\alpha$
and compare them with exact results in quantum mechanics.  First, the
amplitude of two-anyon scattering at the tree level
(Fig.~\ref{fig:aharonov-bohm}) is given by
\begin{equation}\label{eq:amplitude}
 \begin{split}
  A(p,\varphi)
  &=-\frac{4i\pi\alpha}{\sin\varphi}-4{g^*}^2e^{\mp i\varphi} \\
  &=-4i\pi\alpha\frac{e^{\mp2i\varphi}}{\sin\varphi}+O(\alpha^2),
 \end{split}
\end{equation}
where $p$ is the magnitude of the relative momentum and $\varphi$ is the
scattering angle.  The exact scattering amplitude obtained by solving
the Schr\"odinger equation (\ref{eq:schrodinger}) with the boundary
condition (\ref{eq:wavefunction}) at $a\to\infty$ is 
\begin{equation}
 f_\mathrm{exact}(p,\varphi)=\frac{\sin\pi\alpha}{\sqrt{{i\pi p}}}
  \frac{e^{\mp2i\varphi}}{\sin\varphi}
  \qquad\text{for}\quad\varphi\neq0,\,\pi.
\end{equation}
Thus, up to a kinematical factor, the perturbative result
(\ref{eq:amplitude}) reproduces the correct expanded form of the
exact result to the leading order in $\alpha$.

The ground state energy of two anyons in a harmonic potential can be
directly extracted from the scaling dimension of the operator
$\phi$~\cite{Nishida:2007pj}.  At the fixed point, the scaling dimension
of $\phi$ becomes
\begin{equation}
 \Delta_\phi=1+\gamma_\phi=1+|\alpha|+O(\alpha^2),
\end{equation}
from which the ground state energy is given by $E=\Delta_\phi\,\omega$
with $\omega$ being the oscillator frequency.  This result coincides
with the exact result obtained in quantum mechanics,
\begin{equation}
 E_\mathrm{exact}=\left(1+|\alpha|\right)\omega,
\end{equation}
including the energy of center of mass motion.  The corresponding
relative wave function is
\begin{equation}
 \Phi(r,\varphi)=\frac{e^{\mp i\varphi}}{r^{1-|\alpha|}}e^{-\omega r^2/4},
\end{equation}
which satisfies the boundary condition (\ref{eq:wavefunction}) with
$a\to\infty$.

These results establish that the fixed point (\ref{eq:g*}) describes the
physics of anyons with the two-body contact interaction tuned to the
resonance.

\section{RG equations for $v_3$ and $v_4$}
\begin{figure}[tp]
 \includegraphics[width=0.3\textwidth,clip]{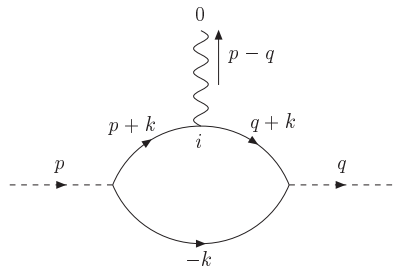}
 \caption{One-loop diagram to renormalize the vertex
 $-i\a\cdot\phi^*\tensor\grad\phi/2$.  This diagram also contributes to
 the renormalization of $v_3$ and $v_4$.  \label{fig:gauge_coupling}}
\end{figure}

Now we study the renormalization of the three-body and four-body
couplings $v_3$ and $v_4$.  First, we shall look at the one-loop diagram
in Fig.~\ref{fig:gauge_coupling}.  Integrating out modes in the momentum
shell $e^{-s}\Lambda<k<\Lambda$, we obtain
\begin{equation}
 i\frac{g^2}{2\pi}\left[(\p+\q)_i
  \mp i\epsilon^{ij}(\p-\q)_j\right]
 \ln\frac\Lambda{e^{-s}\Lambda}.
\end{equation}
The first term renormalizes the vertex
$-i\a\cdot\phi^*\tensor\grad\phi/2$ and is consistent with the
wave-function renormalization of $\phi$.  The second term generates the
following vertex in the Lagrangian density:
\begin{equation}
 \delta\Lag=\mp\frac{g^2}{2\pi}s\,\grad\times\a\,\phi^*\phi.
\end{equation}
Using the CS Gauss law (\ref{eq:gauss}), the generated vertex
$\delta\Lag$ is converted into the renormalization of the couplings
$v_3$ and $v_4$.

\begin{figure}[bp]
 \includegraphics[width=0.48\textwidth,clip]{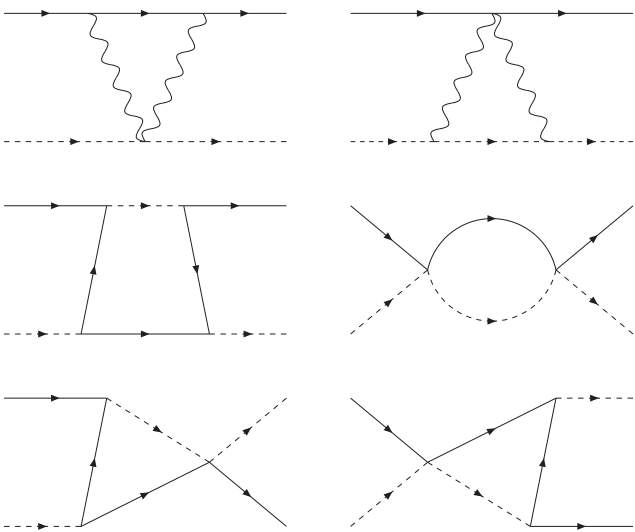}
 \caption{One-loop diagrams to renormalize the three-body coupling
 $v_3$. \label{fig:three-anyon}}
\end{figure}

\begin{figure}[bp]
 \includegraphics[width=0.48\textwidth,clip]{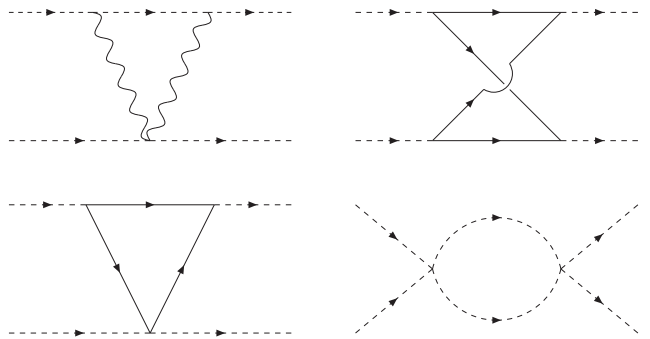}
 \caption{One-loop diagrams to renormalize the four-body coupling
 $v_4$. \label{fig:four-anyon}}
\end{figure}

The other one-loop diagrams that contribute to the renormalization of 
$v_3$ and $v_4$ are depicted in Figs.~\ref{fig:three-anyon} and
\ref{fig:four-anyon}, respectively.  Taking into account the
contribution of the wave-function renormalization of $\phi$, RG
equations that govern the running of $v_3$ and $v_4$ are given by
\begin{align}
 \frac{dv_3}{ds}  \label{eq:v_3}
  %&= |\alpha|g^2 - 6\pi\alpha^2 + \frac{8g^4}{3\pi}
  %- \frac{11g^2}{3\pi}v_3 + \frac{2}{3\pi}v_3^{\,2} \\
  &= \frac{20\pi}{3}\alpha^2 - \frac{22}{3}|\alpha|v_3
  + \frac{2}{3\pi}v_3^{\,2}, \phantom{\frac{}{\int}} \\
 \frac{dv_4}{ds}  \label{eq:v_4}
  %&= 2|\alpha|g^2 - 8\pi\alpha^2 - \frac{4g^4}\pi
  %+ \frac{2g^2}\pi v_3 - \frac{2g^2}\pi v_4 + \frac{2}{\pi}v_4^{\,2} \\
  &= - 20\pi\alpha^2 + 4|\alpha|v_3 - 4|\alpha|v_4 + \frac{2}{\pi}v_4^{\,2}.
\end{align}
Here $g=g^*$ was substituted.  We note that the $\beta$ function of
$v_3$ ($v_4$) is a quadratic function of $v_3$ ($v_4$) to all orders.
Higher order corrections appear only in their coefficients and do not
alter the following discussions as far as the perturbative expansion is
valid $|\alpha|\ll1$.

Equation (\ref{eq:v_3}) has two fixed points at
\begin{equation}
 v_3^*=\pi|\alpha| \qquad\text{and}\qquad v_3^*=10\pi|\alpha|, 
\end{equation}
each of which corresponds to a scale-independent short range boundary
condition on the three-body wave function.  The stable fixed point
($v_3^*=\pi|\alpha|$) corresponds to the regular wave function without
the three-body resonance, while the unstable one ($v_3^*=10\pi|\alpha|$)
corresponds to the three-body resonance at threshold.  When $v_3$ is at
the former fixed point $v_3^*=\pi|\alpha|$, Eq.~(\ref{eq:v_4}) also has
stable and unstable fixed points at
\begin{equation}
 v_4^*=-2\pi|\alpha| \qquad\text{and}\qquad v_4^*=4\pi|\alpha|.
\end{equation}
The unstable fixed point $v_4^*=4\pi|\alpha|$ corresponds to the
four-body resonance at threshold.

However, when $v_3$ is at the other fixed point corresponding to the
three-body resonance, Eq.~(\ref{eq:v_4}) does not have a fixed point.
In this case, the solution of the coupled equations (\ref{eq:v_3}) and
(\ref{eq:v_4}) is 
\begin{equation}\label{eq:limit_cycle}
 \begin{split}
  v_3(s) &= 10\pi|\alpha|, \phantom{\frac{}{}} \\
  v_4(s) &= \pi|\alpha|+3\pi|\alpha|\tan\left(6|\alpha|s+C\right),
 \end{split}
\end{equation}
where $C$ is a constant of integration.  We thus find a RG limit
cycle where $v_3(s)$ is a constant and $v_4(s)$ is a periodic function
of $s$ with a primitive period $\pi/(6|\alpha|)$.  The system tuned to
this limit cycle has a discrete scaling symmetry with a scaling factor
$\exp[\pi/(6|\alpha|)]$ in momentum.  It especially implies a geometric
spectrum of energy eigenvalues~\cite{Mueller04} in the four-anyon system
when two-body and three-body interactions are simultaneously tuned to
the resonance.  When the statistics is close to fermion $|\alpha|\ll1$,
the ratio of two successive energy eigenvalues is given by
\begin{equation}
 \frac{E_{n+1}}{E_n}\to \exp\!\left[-\frac{\pi}{3|\alpha|+O(\alpha^2)}\right].
\end{equation}
There are an infinite number of bound states with an accumulation point
at zero energy.  This resembles Efimov states in three-body systems at
infinite scattering length in three spatial
dimensions~\cite{Efimov70,Albeverio:1981zi,Bedaque:1998kg}, whose
experimental evidence has been recently reported using cold atoms in
Ref.~\cite{Kraemer06}.

Near the bosonic limit, on the other hand, the Efimov states do not
exist in a few-anyon system.  This is because the two-body resonant
interaction becomes just zero interaction in the bosonic limit.
Therefore, there should be at least one critical value of $\alpha$ where
the four-anyon bound states disappear.  Determination of such a critical
$\alpha$ is, however, beyond the scope of our perturbative approach.

\begin{figure}[tp]
 \includegraphics[width=0.46\textwidth,clip]{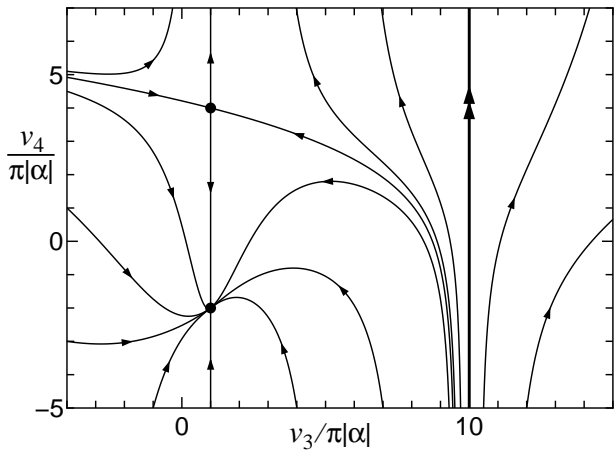}
 \caption{RG flow diagram in the plane of $v_3$ and $v_4$ at $g=g^*$.
 There are two fixed points (dots) and one limit cycle (thick line with
 two arrows).  The arrows indicate flows toward the infrared limit.
 \label{fig:RG_flow}}
\end{figure}

The RG flow diagram obtained from Eqs.~(\ref{eq:v_3}) and (\ref{eq:v_4})
is shown in Fig.~\ref{fig:RG_flow}.  We see two fixed points located at
\begin{equation}
 (v_3^*,v_4^*)=(\pi|\alpha|,-2\pi|\alpha|)
  \quad\text{and}\quad (\pi|\alpha|,4\pi|\alpha|), 
\end{equation}
and the limit cycle given by the solution (\ref{eq:limit_cycle}).

\section{Scaling dimensions of operators}
It is worthwhile to determine scaling dimensions of $N$-body
operators at the fixed points because they are observable as energy
eigenvalues in a harmonic potential~\cite{Nishida:2007pj}.  At the
one-loop level, anomalous dimensions of three-body and four-body
composite operators $\phi\psi$ and $\phi\phi$ are, respectively,
computed as
\begin{equation}
 \begin{split}
  \gamma_{\phi\psi} &= \frac{8|\alpha|}3-\frac{2v_3^*}{3\pi}
  +O(\alpha^2),\\
  \gamma_{\phi\phi} &= -\frac{2v_4^*}{\pi} +O(\alpha^2).
 \end{split}
\end{equation}
From those, we obtain the scaling dimension of the lowest $(2n+1)$-body
operator $\phi^n\psi$ as
\begin{equation}
 \Delta_{\phi^n\psi}=n+1+n\gamma_\phi+n\gamma_{\phi\psi}
\end{equation}
and that of the lowest $(2n)$-body operator $\phi^n$ as
\begin{equation}
 \Delta_{\phi^n}=n+n\gamma_\phi+\frac{n(n-1)}2\gamma_{\phi\phi}
\end{equation}
up to the order $\alpha$.  Then the ground state energy of $N$ anyons in
a harmonic potential is given by the scaling dimension of the $N$-body
operator times $\omega$.  The leading-order results can be easily
understood by recalling that, in the fermionic limit $|\alpha|\to0$,
fermion pairs at the $p$-wave resonance form point-like bosons and they
do not interact with each other or with extra fermions.  Since these
particles can occupy the same state, the ground state energy is simply
given by the number of particles multiplied by the lowest single
particle energy $\omega$ in a harmonic potential.

\section{Many-body physics}
An interesting extension of our work is to the many-body system of
anyons.  Here we discuss some qualitative features of such a system.
When the two-body resonant interaction is present, we have choices to
further introduce the three-body and four-body resonant interactions.
If the three-body resonant interaction is introduced, the Efimov effect
occurs in the four-anyon system (the limit cycle in
Fig.~\ref{fig:RG_flow}) and accordingly the corresponding many-body
system can not be stable.  Instead, if the four-body resonant
interaction is introduced (the upper fixed point in
Fig.~\ref{fig:RG_flow}), the corresponding many-body system will be
unstable because of the attractive interaction between bosonic dimers. 
Therefore the only stable many-body system seems to be the case where
the two-body resonant interaction is present without the three-body and
four-body resonant interactions (the lower fixed point in
Fig.~\ref{fig:RG_flow}).

In this case, we define $\xi$ as a ratio of the ground state energy $E$
to that of ``noninteracting'' anyons $E_0$ at the same density and
statistics; $\xi=E/E_0|_{n,\alpha}$.  Since the system has no intrinsic
dimensionful quantity, $\xi$ can depend only on the statistics parameter
$\alpha$.  Although it is a difficult many-body problem to determine
$\xi$ in general $\alpha$, the problem becomes treatable near the
fermionic limit.  In the fermionic limit, the ground state energy $E$
vanishes because resonantly interacting anyons form noninteracting
point-like bosons, while $E_0$ remains nonzero because of the Fermi
energy.  Therefore we have $\xi\to0$ in the fermionic limit
$\alpha\to0$.  Corrections to $\xi$ near the fermionic limit are
computable using the perturbative framework developed in this article.
It will be important in future works to further study many-body
properties of resonantly interacting anyons, for example, whether they
exhibit superfluidity, and their implications for real planar systems.

\section{Conclusion}
We have investigated anyons with a contact interaction by formulating a
field theory that becomes perturbative near the fermionic limit.  We
identified a fixed point that describes anyons at the two-body
resonance.  In RG equations for three-body and four-body couplings, we
found two nontrivial fixed points and computed scaling dimensions of
$N$-body operators which are observable as a ground state energy of $N$
anyons in a harmonic potential. 

Furthermore, we found a limit cycle behavior in the four-body coupling,
which implies an infinite set of discrete bound states (Efimov states)
in the four-anyon system when both the two-body and three-body
interactions are tuned to the resonance.  Such a system provides a
unique opportunity to study Efimov-type physics within a perturbative
framework.  It should be possible to verify these results by solving a
few-body Schr\"odinger equation for anyons with suitable short range
boundary conditions and hopefully by experiments in the future.  We
believe this work opens up new universal physics in the system of anyons
in two spatial dimensions.

\acknowledgments
The author thanks D.~T.~Son for useful discussions.  This work was
supported by JSPS Postdoctoral Program for Research Abroad.

\end{document}